\documentclass[prb,preprint]{revtex4-1}

\usepackage{amsmath}  
\usepackage{amsfonts} 
\usepackage{graphicx} 
\usepackage{soul} 
\usepackage{xcolor} 
\usepackage{mathrsfs} 

\begin{document}

\title{Measuring optical force with a torsion pendulum: a platform for independent student experimentation}

\author{Leland Russell}
\author{Ezekiel A. Rein}
\author{Anatalya Piatigorsky}
\author{Jennifer T. Heath}
\email{jheath@reed.edu}

\affiliation{Department of Physics, Reed College, Portland OR 97202-8199}

\date{\today}


\begin{abstract}
In this work, the force due to radiation pressure is measured with sub-10 pN sensitivity, corresponding to less than 2 mW of optical power. The apparatus adds homemade reflectors to a commercial Cavendish balance, which consists of a torsion pendulum with a built-in capacitance position sensor. When driven by four 5 mW laser diodes, with square-wave modulation at the pendulum’s natural frequency, the response is strong enough to easily discern in a short time series. The discrete Fourier transform of a longer dataset provides a more in-depth analysis, clearly showing the multiple frequency components from the square-wave driving force. The driving power was controlled by adjusting the square wave duty cycle, allowing easy automation and avoiding additional optics or filters. For a 9-hour dataset, white noise corresponding to about 2 pN was observed, enabling our most sensitive measurements. The pendulum  operates in air. To minimize convective forces from differential heating and the resulting differential pressure, we use symmetrical reflectors encased in low-thermal conductivity material, namely, two glass-fronted mirrors attached back-to-back. This experiment could be used in a single lab session, allowing the optical  force exerted by a laser pointer to be quickly and intuitively observed. It also demonstrates the power of Fourier analysis, builds student intuition about oscillator systems, and provides a compelling platform for student-driven projects. 
\end{abstract}

\maketitle 

\section{Introduction} 

Radiation pressure plays a key role across a wide range of physics research topics, from astrophysical models \cite{1} to laser cooling \cite{2} and optical tweezers,\cite{3} and has been demonstrated in undergraduate projects, including, recently, optical levitation and manipulation of a silicon oil droplet.\cite{4} This present work was motivated by a fascination with the gigantic solar sails being designed to propel space missions using radiation pressure.\cite{5} An early test satellite was launched by Japan in 2010,\cite{6} and NASA’s Advanced Composite Solar Sail mission (ACS3) launched in April 2024.\cite{7} The fundamentals of radiation pressure, including momentum transfer across dielectric surfaces,\cite{8} continue to be a topic of research interest. Despite clear evidence of radiation pressure, the idea that a massless particle can truly exert a force is somewhat counterintuitive to students. 

In the experiment described here, the force exerted by optical radiation, henceforth called the optical  force, drives a torsion pendulum. The driving force can be quantitatively measured through analysis of the pendulum’s motion. This enables student projects measuring optical  force, or alternatively, projects that explore the impact of an arbitrary driving force on a harmonic oscillator.

While it’s common for students to encounter driven and damped oscillator systems in an electronics context, it can be helpful to see mechanical systems through which such topics can be studied. Torsion pendulums are used widely in experimental physics research for sensitive force measurements. Current state-of-the-art measurements achieve sub-femtoNewton (fN) sensitivity, taking advantage of the dramatic increase in quality factor $Q$ that results from placing the pendulum in vacuum.\cite{10} This approach enables studies of the  optical  force itself, as well as the equivalence principle between gravitational and inertial mass,\cite{11,12} the Casimir effect,\cite{13} and fundamental sources of noise.\cite{14} Pendulum systems are also being developed for space-based gravitational wave observatories\cite{15} and other large experiments. Sometimes  optical  forces are even used as part of a feedback system to damp unwanted vibrations of pendulum systems.\cite{10} Of course, oscillator systems in general play a role throughout physics, and the idea that different frequency channels can carry different information is key to a wide range of experimental systems from optics and electronics to probe microscopy. 

The first experimental measurements of  optical  force were carried out independently by two separate groups: Peter Lebedev from University of Moscow (1899)\cite{16}, and Gordon F. Nichols and Earnest F. Hull (1901),\cite{17,18} at Dartmouth College in the United States. Early measurements were impressive feats of experimental design, and in addition to their scientific merits, they make a useful case study for discussion of scientific method and ethics. Apparently neither coauthors nor contemporary readers checked the work too carefully, perhaps because the results agreed with theoretical predictions. Years later, the Nichols and Hull work was found to contain significant errors,\cite{19} including confusing a base-e logarithm with a base-10 logarithm and incorrectly applying their galvanometer calibration factor in the data analysis. The larger story, as explained by Elsa Garmire,\cite{20} highlights the importance of careful peer-review, the danger of confirmation bias,\cite{21} and the complexity of assigning attribution for scientific discovery. 

Much of the difficulty in these historic measurements arose from trying to distinguish the force due to radiation pressure from that due to differential heating, and hence differences in gas pressure (also called convection forces). To allow for a simple setup with a commercial pendulum, the apparatus described here does operate at atmospheric pressure and is thus vulnerable to this same concern. 

The discussion below starts with a brief overview of the optical force, and the response of a torsion pendulum to both constant and time-dependent torques. We then describe our experimental setup and demonstrate its sensitivity. 

\section{Theory}
\subsection{Optical force}

Light, being an electromagnetic wave, carries both energy and momentum. The flow of energy is described by the Poynting vector $\langle \vec{S} \rangle$, in the sense that energy carried by the electromagnetic fields across a surface of area $A_{\mathrm{surf}}$ with normal $\hat{n}$ in time $\Delta t$ is given by $\vec{S} \cdot (A_{\mathrm{surf}} \, \hat{n})\Delta t$. Then, the average power (energy per time) incident on an evenly illuminated surface is 
\begin{equation}
P=\langle S \rangle A_{\mathrm{surf}}\cos \theta \hspace{5mm} \mathrm{(fully\,\,illuminated\,\,sample)},
\label{illum}
\end{equation}
where $\theta$ is the angle of incidence of the beam. This expression is relevant when the illumination is spatially uniform over the entire surface of the target, as is the case for a solar sail. For our experiment, the laser-illuminated spot is smaller than the target surface, giving instead the relation
\begin{equation}
P=\langle S \rangle A_{\mathrm{beam}} \hspace{5mm} \mathrm{(small\,\,beam)}.
\label{P&S}
\end{equation}
where $A_{\mathrm{beam}}$ is the cross-sectional area of the beam, which is smaller than the spot size on the surface by a factor of $\cos\theta$.

The Poynting vector is related to the momentum density in the fields $\vec{g}$ (momentum per volume) by the speed of light $c$ as \cite{Griffiths} 
\begin{equation}
\frac{\vec{S}}{c^2}=\vec{g}.
\label{S&g}
\end{equation}
If we wait for a time $\Delta t$, the total momentum crossing a surface perpendicular to the beam will be
\begin{equation}
\Delta \vec{p}=\langle \vec{g} \rangle c \, \Delta t \, A_{\mathrm{beam}}
\label{p&g}
\end{equation}
since a volume with cross sectional area $A_{\mathrm{beam}}$ and length $c \Delta t$ of our electromagnetic
wave passes through the area $A_{\mathrm{beam}}$ in time $\Delta t$. If the light is absorbed by the
surface, this momentum imparts a force to the surface given by $\vec{F}=\Delta \vec{p}/{\Delta t}$. Using Eq. (\ref{P&S}), (\ref{S&g}), and (\ref{p&g}), and considering only the component of force perpendicular to the target surface, $F_\perp$ yields 
\begin{equation}
F_\perp=\frac{P}{c}\cos \theta  \hspace{5mm}\mathrm{(complete\,absorption,\, small\, beam)}
\label{Fperp}
\end{equation}
Optical emission follows a similar process, thus for reflective surfaces, the total force perpendicular to the surface is 
\begin{equation}
F_\perp = (\vec{F_i}-\vec{F_r})\cdot\hat{n}=\frac{P_i}{c}(1+\alpha)\cos\theta \hspace{5mm}\mathrm{(partial\,reflection, \, small\, beam)}
\end{equation}
where $\vec{F_i}$ and $\vec{F_r}$ are the forces from incident and reflected light, respectively. The reflectivity, $\alpha = P_r/P_i$, is often approximated as 1 for highly reflective surfaces; $P_i$ and $P_r$ are the incident and reflected optical power. 

In the case of a fully illuminated sample, these expressions are typically written in terms of the optical energy flux density $\mathscr W$ (power per area) and the radiation pressure $\mathscr{P}$ (perpendicular force per area).\cite{20b} Using a similar approach starting from Eq. (\ref{illum}) yields:
\begin{equation}
\mathscr P=\frac{\mathscr W}{c}\cos^2 \theta \, (1+\alpha) \hspace{5mm} \mathrm{(fully\,\,illuminated\,\,sample)}.
\end{equation}

When $\theta=0$, as it is in this experiment, the relationship between optical power and the magnitude of the force is remarkably simple:  
\begin{equation}
F=(1+\alpha)\frac{P}{c}
\label{optF}
\end{equation}
The final result, that the radiation force (momentum per time) given by Eq. (\ref{optF}) is simply $1/c$ times the power (energy per time) is to be expected from the familiar energy-momentum relation for a photon $E_{\gamma}=p_{\gamma} c$.

If, for example, one is using two 5 mW laser pointers for a total of 10 mW of optical power, and the mirror surface is close to perfectly reflecting, then the force is $6.7\times 10^{-11}$N . Thus, for this experiment we need to be able to accurately measure forces on the scale of tens of pN. 

\subsection{Pendulum response with constant torque}

\begin{figure}[h!]
\centering
\includegraphics[width=3in]{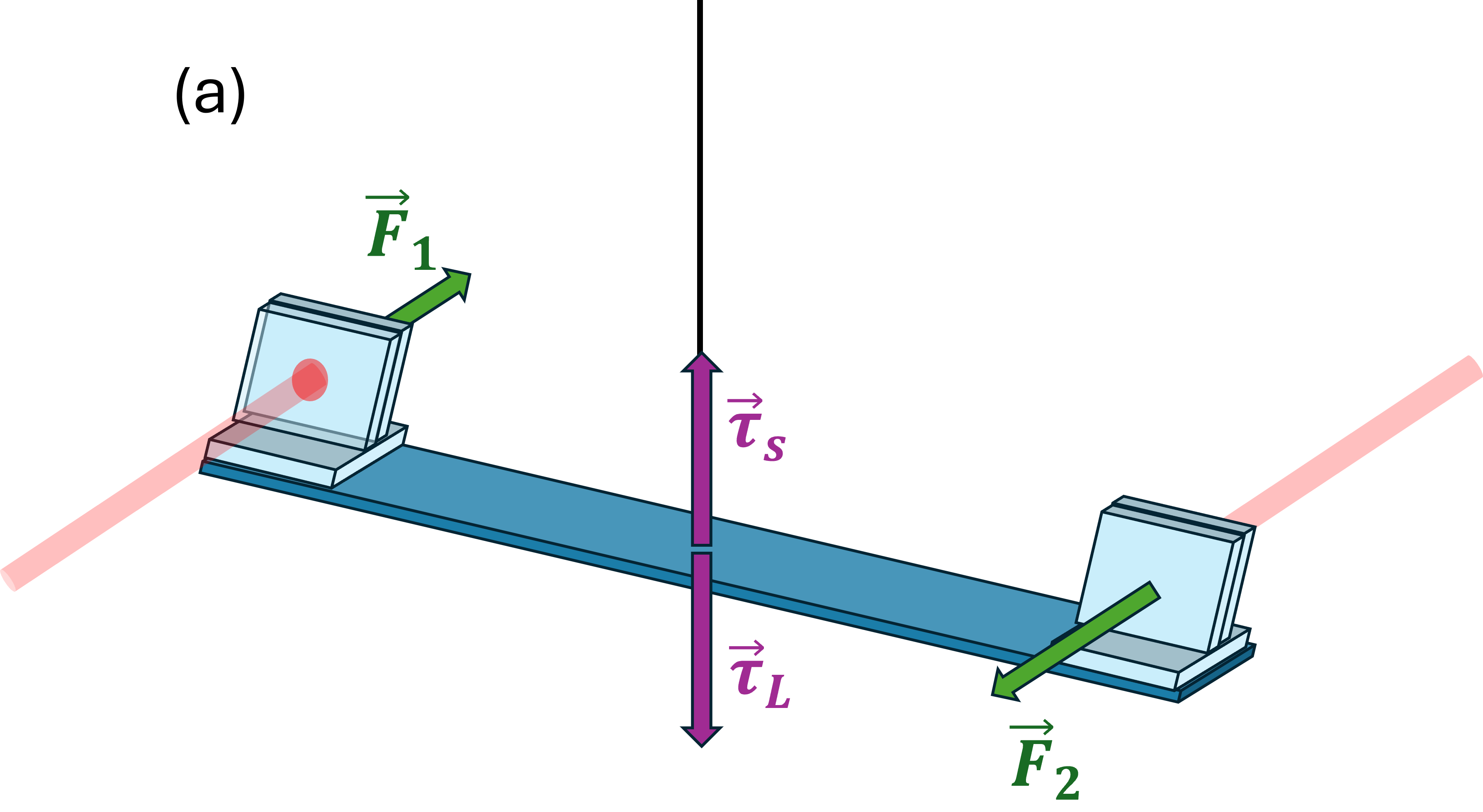}
\includegraphics[width=3in]{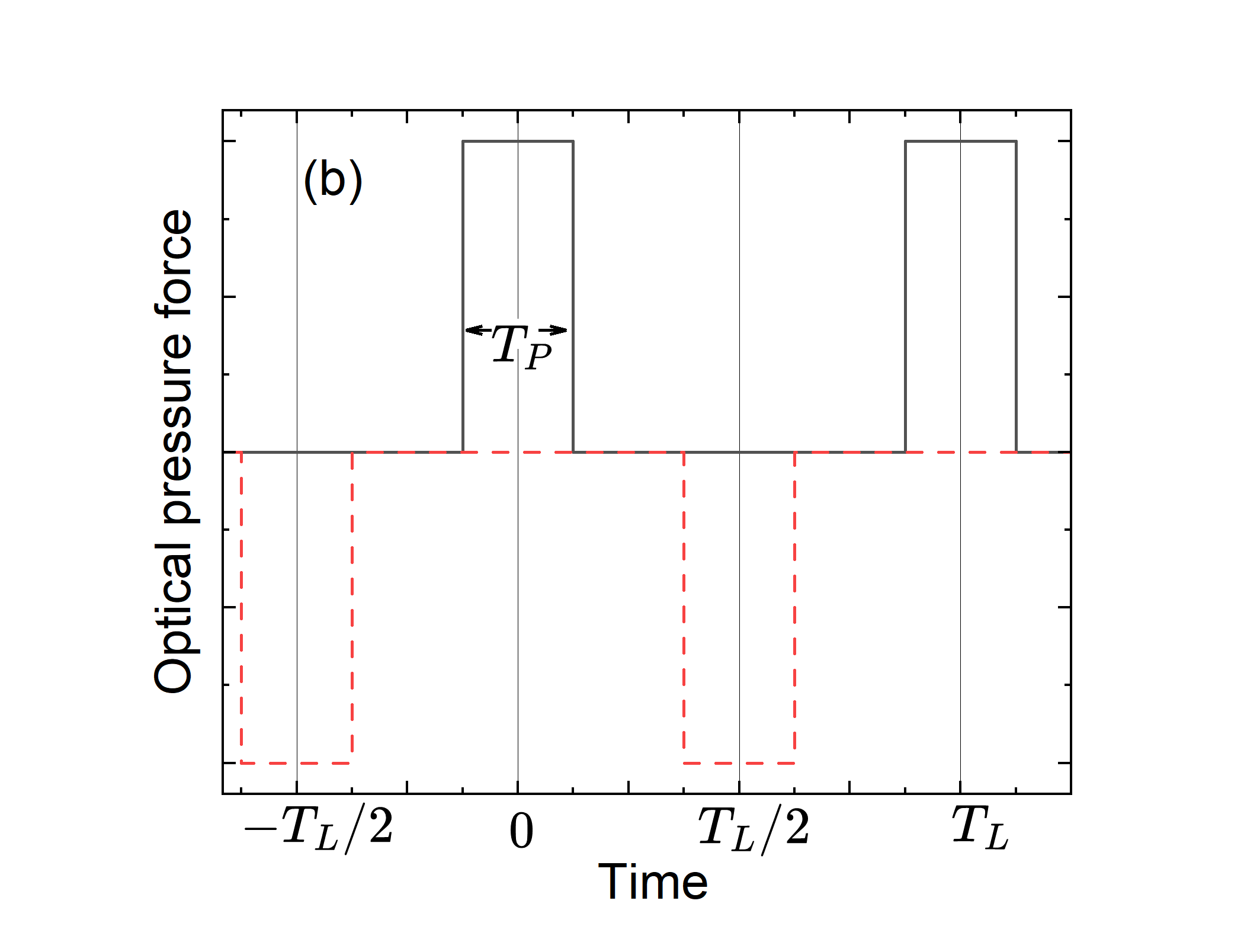}
\caption{\textbf{(a)} Schematic illustrating forces and torques on the pendulum. \textbf{(b)} Laser pulse timing. Two lasers could combine to give the timing shown (solid line), with duty cycle $d=T_P⁄T_L$. To increase the forcing, two additional lasers could be aligned opposite of those shown in (a), with pulsing offset by $T_L⁄2$ (dashed line).}
\label{schematic}
\end{figure}

The torsion pendulum consists of a horizontal boom hanging from a thin wire as shown in Fig. \ref{schematic}. An external torque $\vec{\tau}$ could be created by  exerting  optical  forces $\vec{F_1}$ and $\vec{F_2}$ at the ends of the boom, yielding the optical torque $\vec{\tau_L}=(\vec{r_1}\times\vec{F_1})+(\vec{r_2}\times\vec{F_2})$ where $\vec{r_1}$ and $\vec{r_2}$ point from the wire to each laser spot. When twisted, the suspending wire exerts a torque response $\vec{\tau_S}$, related to its spring constant $\kappa$ and the twist angle $\vec{\phi}$ as $\vec{\tau_S}=-\kappa\vec{\phi}$. 
 	
Since the relevant angles and torques will all point along one axis (aligned with the suspension wire), we will hereafter drop the vector notation and simply use positive or negative values to indicate directions. In static equilibrium, the angular acceleration is zero, hence
\begin{equation}
    \tau_L=\kappa\phi.
    \label{DC}
\end{equation}
\subsection{Dynamic pendulum response}

Inevitably, the pendulum is not in static equilibrium but has been set swinging by noise, vibration, or changes in optical power. The motion is also continuously damped. We expect the energy dissipation to be dominated by viscous damping of the surrounding air. However, interactions in the capacitive position sensor may also play a role. We do not expect significant damping by the suspending wire, though in vacuum systems this can dominate. \cite{20c} Using a simple model, viscous damping is related to the angular velocity $\dot{\phi}$ by a proportionality constant $b$. Identifying the natural frequency of oscillation $\omega_0=2\pi f_0 = \sqrt{\kappa /I}$, and using $\gamma=b/I$ yields a differential equation modeling the pendulum’s motion:
\begin{equation}
\ddot{\phi}+\gamma\dot{\phi}+{\omega_0}^2 \phi = \frac{\tau(t)}{I}.
\label{diffeq}
\end{equation}
where $I$ is the moment of inertia. 

Under a constant torque $\tau_0$,  the pendulum oscillates around an equilibrium angle $\phi_0=\tau_0/\kappa$ as \cite{22,23} 
\begin{equation}
\phi(t)=\frac{\tau_0}{\kappa} + A e^{-\gamma t/2} e^{i \omega_1 t}
\end{equation}
with
\begin{equation}
\omega_1=\sqrt{{\omega_0}^2-\gamma^2/4}
\end{equation}
In this work, $\omega_0$ is only about 1\% larger than $\omega_1$, and so we approximate $\omega_1\approx\omega_0$.

For an arbitrary time-dependent excitation, $\tau(t)$, the Fourier transform can be used to evaluate Eq. (\ref{diffeq}). We write the Fourier transform of $\phi(t)$ as
\begin{equation}
\tilde{\phi}(\omega)=\int_{-\infty}^{\infty} \phi(t) e^{-i\omega t}\,dt
\label{FT}
\end{equation}
and the inverse Fourier transform as
\begin{equation}
\phi(t)=\frac{1}{2\pi}\int_{-\infty}^{\infty} \tilde{\phi}(\omega) e^{i\omega t}\,d\omega.
\label{invFT}
\end{equation}
Equation (\ref{invFT}), along with the analogous expression for $\tau(t)$, can be substituted into Eq. (\ref{diffeq}) to yield:
\begin{equation}
\frac{1}{2 \pi} \left( \frac{d^2}{dt^2} + \gamma \frac{d}{dt} + {\omega_0}^2 \right) \int_{-\infty}^{\infty} \tilde{\phi}(\omega)e^{i\omega t} d\omega=\frac{1}{2\pi}\int_{-\infty}^{\infty} \frac{\tilde{\tau}(\omega)}{I} e^{i\omega t} d\omega.
\end{equation}
Switching the order of operations on the left leads to the solution
\begin{equation}
    \tilde{\phi}(\omega)=\frac{\tilde{\tau}(\omega)}{I({\omega_0}^2-\omega^2+i\gamma \omega)}
\end{equation}
The pendulum will oscillate at the driving frequency (or frequencies) with amplitude
\begin{equation}
    |\tilde{\phi}(\omega)|=B(\omega)\,|\tilde{\tau}(\omega)|,
    \label{phiBtau}
\end{equation}
where $B(\omega)$ is the pendulum response function
\begin{equation}
B( \omega )=\frac{1}{I \sqrt{({\omega_0}^2-\omega^2)^2+\gamma^2 \omega^2 }}.
\label{B}
\end{equation}

As we think about the pendulum behavior in general and try to optimize the experimental design, it can be helpful to cast $B(\omega)$ in terms of the unitless quantity $x=\omega/\omega_0$ and write Eq. (\ref{B}) as
\begin{equation}
B( x )=\frac{1}{\kappa \sqrt{(1-x^2)^2+{(x/Q)}^2 }}
\end{equation}
where $Q=\omega_0/\gamma$ is the quality factor. 

The most sensitive AC measurements can be made when driving at $\omega \approx \omega_0$, which gives $B=Q⁄\kappa$. When $\omega \ll \omega_0$, then $x\ll 1$ and  $B=1⁄\kappa$, a constant value; this is the steady-state behavior of the pendulum, Eq. (\ref{DC}). For $\omega \gg \omega_0$, $x \gg 1$ and $B$ is proportional to $\omega^{-2}$, approaching zero for large frequencies. 

Thus, for an experiment using modulated light, it’s useful to have a lightweight pendulum with a high quality factor, and to modulate the light near the natural resonance frequency. 
Alternatively, to see a constant force, a pendulum with a larger value of $I$ could be useful, as the pendulum will not respond as significantly to various sources of noise, and its position will seem more stable. The deflection at zero frequency does not depend on $I$, except to the extent that a greater tension on the suspending wire may cause $\kappa$ to vary slightly.

\subsection{Dynamic driving force}

Suppose the forcing function $F_L(t)$ is a periodic pulse with amplitude $F_0$, offset $C$, and pulse width $T_p$ resulting in a duty cycle $d=T_p⁄T_L$  where $T_L$ is the period. This function, illustrated in Fig. \ref{schematic}b, can be defined over one period as
\begin{equation}
F_L(t) =
\begin{cases}
	C+F_0 & |t| \le T_p/2 \, , \\
	C &  T_p/2 < |t| \le T_L/2 \, .
\end{cases}
\end{equation}

Assuming the pendulum is driven for a time much longer than $T_L$, the Fourier series representation gives the frequency dependence of this function:	
\begin{equation}
F_L(t)=C+dF_0+\sum_{n=1}^\infty F_n \cos(n \omega_L t)
\end{equation}
with $\omega_L=2 \pi / T_L$ and 
\begin{equation}
F_n=\frac{2}{n \pi} F_0 \sin(n \pi d).
\label{FTsq}
\end{equation}
For a symmetric square wave, $d=0.5$ and components with even $n$ are zero. The fundamental component is 
\begin{equation}
F_1=\frac{2}{\pi}F_0 \hspace{5mm} (d=0.5)
\end{equation}

The pendulum will also be constantly driven by noise and vibrations. Even in a perfectly quiet environment, the noise floor of the measurement would be limited because the viscous interactions that dissipate energy also cause white noise (Brownian motion). As described by the fluctuation-dissipation theorem, this creates a frequency-independent torque on the pendulum, resulting in deflections whose amplitude depend only on the pendulum response function $B$:
\begin{equation}
\tilde{\phi}_B(f)=\sqrt{2 \langle \phi^2(f) \rangle}=B(f)\sqrt{8k_BTI\gamma \Delta f}
\label{Bnoise}
\end{equation}
where $\tilde{\phi}_B(f)$ is the predicted Brownian component of the motion and $\Delta f$ is the bandwidth, which is typically taken to be the inverse of the record length for the measurement.\cite{24,25} We’ve written Eq. (\ref{Bnoise}) explicitly in terms of $f$ because that is more directly observed parameter in the experiment. It also contains an extra factor of $\sqrt{2}$ in order to express deflection in terms of peak amplitudes, rather than in rms units.
\section{Experimental Methods}
\subsection{The torsion balance}
\begin{figure}[tbh]
    \begin{minipage}{0.48\textwidth}
            \includegraphics[width=\linewidth]{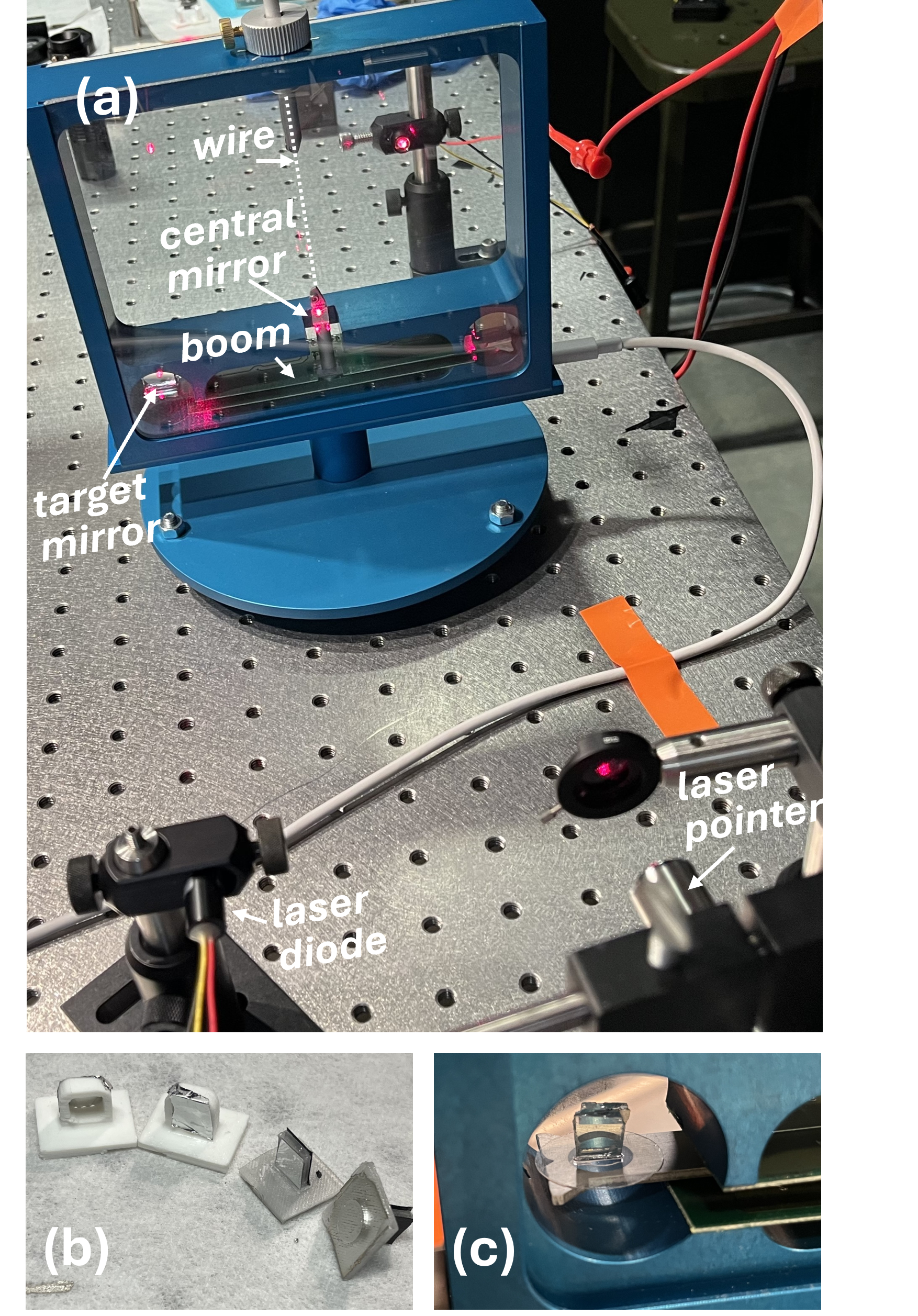}
    \end{minipage}%
    \hspace{\fill}%
    \begin{minipage}{0.48\linewidth}
        \caption{\textbf{(a)} Pendulum setup with key components labeled. The suspending wire is not visible; its location is marked with a dashed line. Normally, the laser pointer (for calibration) and laser diodes (providing driving torque) are not turned on simultaneously. \textbf{(b)} Alternative reflectors. (L-R) back and front of the foil reflector, front and base of the Mylar reflector showing the alignment tab. \textbf{(c)} Glass reflector on the pendulum boom, aligned over the hole in the boom where a lead ball would normally sit.}
        \label{setup}
    \end{minipage}
\end{figure}

This experiment is based on use of the Tel-Atomic Cavendish Balance (TEL-RP2111, \$2600)\cite{9} which includes a capacitance sensor and computerized data collection.\cite{9b} The setup is shown in Fig. \ref{setup}. 
This pendulum has a horizontal aluminum ``boom'' suspended by a 25 $\mu$m diameter tungsten wire. The boom has circular cutouts at each end in which small lead balls can be placed. The boom is inside an enclosed housing, reducing air currents and providing a hard stop maximum rotation of 70 mrad in each direction. The pendulum setup instructions, its parameters relevant to operation and moment of inertia, and the software instructions are provided by Tel-Atomic.\cite{9} In this experiment, the balance is modified by replacing a pair of small lead balls with reflective targets, which is easily reversible.

The pendulum was set up on an optical table that is designed to float, although it was not actually floating during these experiments. The room is small and separated from other laboratories, on the second floor of our building. Environmental vibration did not seem to be an issue for us, as long as no-one physically bumped the table. Room temperature variations did create drift in the sensor reading, especially apparent in summer in our non-air-conditioned laboratory.\cite{9c}

The capacitive position sensor can be calibrated by swinging the pendulum so it is firmly against each hard stop. A calibration screen in the supplied software allows you to indicate a ‘maximum’ and ‘minimum’ deflection (of $\pm$70 mrad). For more accuracy, a central mirror allows calibration by the standard optical lever approach, described in the manufacturer instructions. 
\subsection{Light sources}
While any light source can create a force, we used inexpensive laser diodes (Adafruit P/N 1056 \$19) that could be controlled using a TTL signal. Their 5 mW output was verified using a Newport 818-UV power meter at the start and end of the experiment. This power meter was also used to measure the transmission of the enclosure’s glass window and the specular reflectivity of the mirrors.

We experimented with both a constant (DC)  optical  force and a square-wave modulated (AC) driving force. For a DC force, up to two lasers can be used, as illustrated in Fig. \ref{schematic}. In the AC experiment, this increases to four lasers by using double-sided mirrors and driving the pendulum alternately in positive and negative directions. With twice the light and $Q=4.5$, the AC experiment resulted in almost an order-of-magnitude greater deflection than the DC approach, which transformed this experiment from one in which we struggled to separate the signal from the noise (especially when thermal drift was an issue),\cite{AP} to one in which the signal was clearly evident. Thus, the AC results are the focus of this paper.

In some experiments, the light output was modulated using a function generator. Alternatively, using an Arduino Uno, a series of measurements could run sequentially, collecting data at a range of optical powers. The Arduino was also helpful when using four lasers, as they had to cycle opposite each other, providing both positive and negative torques. The optical forcing could be varied by changing the duty cycle of the square wave, or by changing the number of lasers being flashed from 1 up to 4; the radiation pressure is small enough that the force does not have to be exactly balanced across the ends of the torsion pendulum. We also experimented with using neutral density filters to vary the light intensity, but since we did not have an automated filter wheel, these had to be changed out by hand which was time consuming and caused vibrational noise.
\subsection{Mirrored targets}
In place of the small lead balls of the Cavendish experiment, mirrored targets were balanced on each end of the boom. They simply rested on the boom and were not affixed to it. Most of the data reported here was collected using second surface mirrors created from 0.5” square, 0.063” thick glass-fronted “craft” mirrors (Amazon, \$17.50/25 pieces). These were cleaved to create 0.25” square mirrors. Each target was made of two such squares, attached back-to-back with silver paint. They were epoxied to a glass coverslip that formed a base, as shown in Fig. \ref{setup}. These were then centered and aligned by hand on the horizontal boom of the pendulum. This design resulted in $Q=4.5$ and $I=(1.6\pm 0.1)\times 10^{-5}\, \mathrm{kg\, m^2}$. The value of $I$ was dominated by the boom ($I_{\mathrm{boom}}=(1.21\pm0.08)\times10^{-5}\, \mathrm{kg\, m^2}$, as given in the instrument datasheet).  

By placing the mirrors back-to-back with the reflective surface at the center and connecting them thermally with silver paint, we hope that any heating happens equally at the front and back of the target, thus providing zero net torque. As our results will show, the pendulum response is consistent with optical forcing, so we believe this design is successful.

An alternative mirror target was made by 3D printing a structure and attaching Mylar (Amazon-Vivosun, \$28/100$\mathrm{ft^2}$) with double-sided tape. This created a larger reflective surface that was easier to work with, and that is more similar to the reflectors used in solar sails. The design includes an alignment tab at the bottom that slots into the existing holes on the boom. Like glass, Mylar is an electrical and thermal insulator, which should help reduce the thermally induced forces, and the larger substrate might also help stabilize temperatures. This target unfortunately yielded measurements about three times larger than the theoretically predicted values, suggesting that the symmetric design and/or thicker thermally insulating coating of the glass targets was critical. 

To intentionally explore convection forces, we painted the 3D printed target black. Finally, we also created a 3D printed ‘frame’ structure to allow a suspended metal foil to be used as a target. With either of these targets, the pendulum showed dramatic response to optical illumination, about an order of magnitude greater than the expected  optical  force. The foil target will be briefly discussed below.

Vector files for the 3D printed mounts are provided in the supplemental materials. Although these alternative mounts did not give good measurements of  optical  force, they would be an outstanding choice for a project focusing on the time and frequency response of the pendulum to an arbitrary waveform, as the response is so much larger. 
\subsection{Data collection}
\begin{figure}[h!]
\centering
\includegraphics[width=3in]{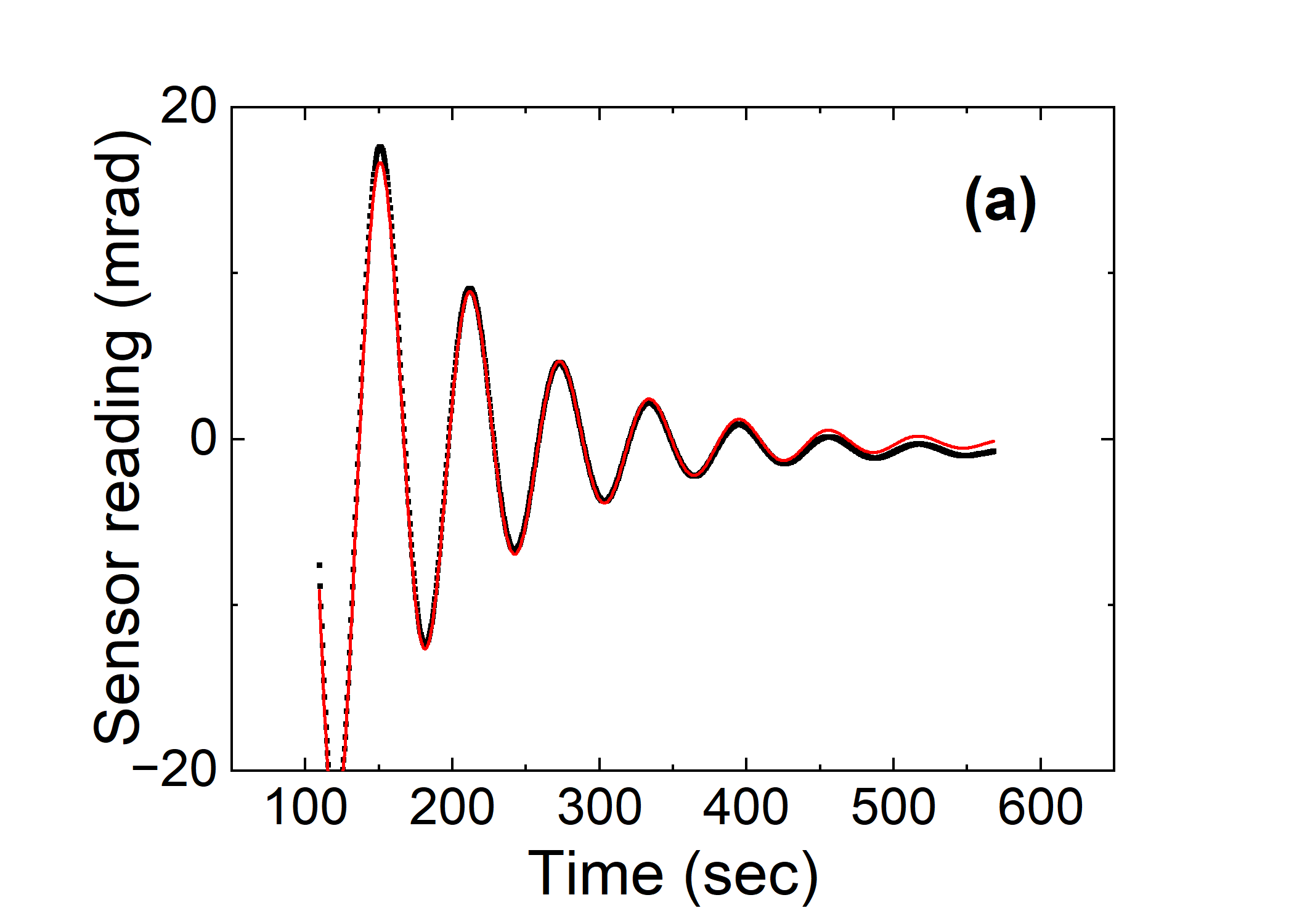}
\includegraphics[width=3in]{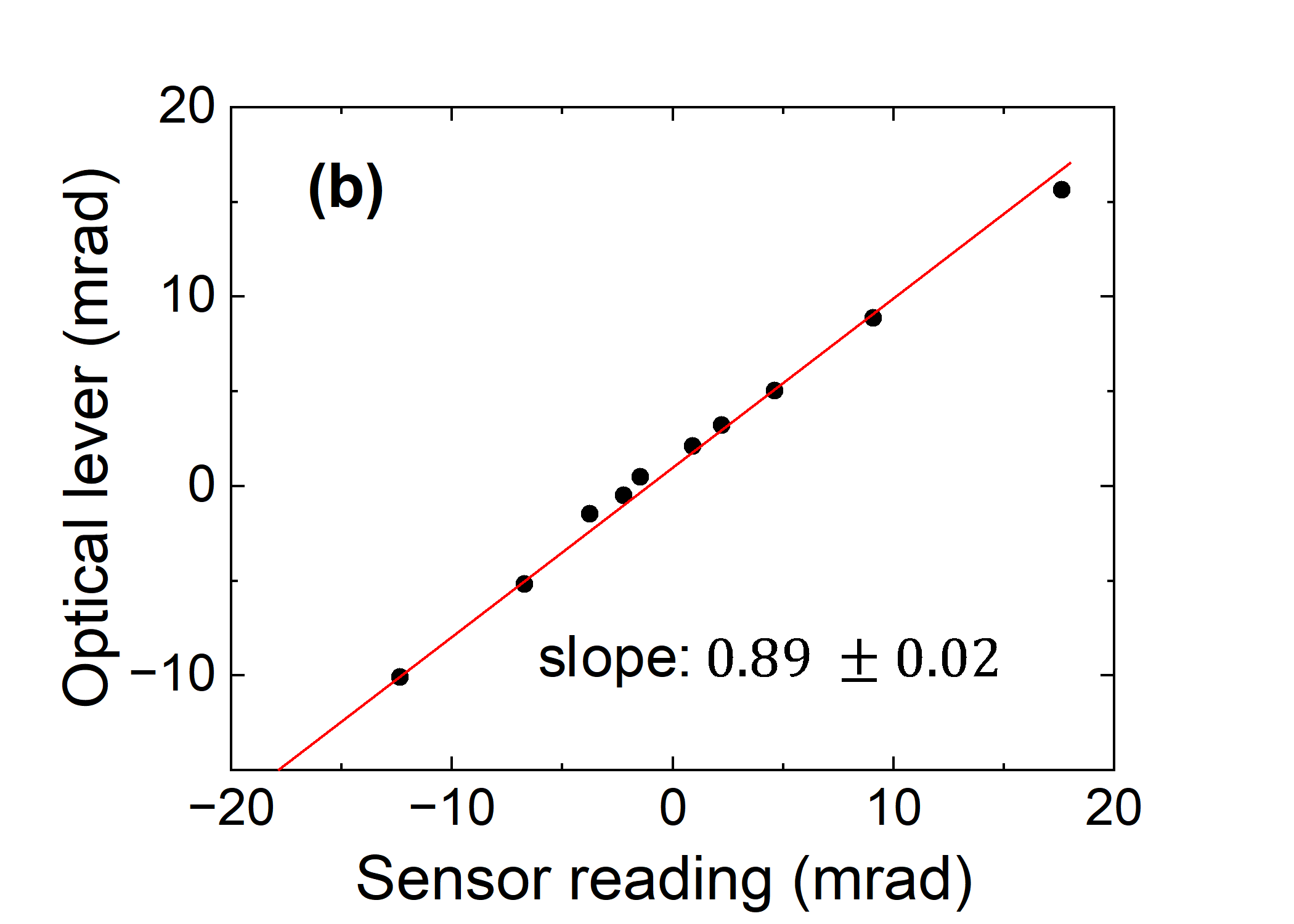}
\caption{\textbf{(a)} Oscillation of the pendulum after a disturbance, fit to a damped sine wave with $f=0.016$ Hz and $\gamma$=0.02 Hz. \textbf{(b)} Optical lever calibration resulting in a correction factor of 0.89. Sensor readings are from the data shown in (a).}
\label{optlever}
\end{figure}

A typical data run started with calibrating the pendulum using the optical lever approach. The pendulum was set up so that a laser pointer reflected off the central mirror hit a piece of paper on the wall about 4 m away. The pendulum was then slightly disturbed to give a damped oscillatory response. While it slowly oscillated (period $\sim 90$ s), the extrema were marked by hand. A graph of these optical lever deflections versus the electronically measured values yields the optical lever calibration factor, as shown in Fig. \ref{optlever}. The electronic data were also fit to a damped oscillator function to find $f_0$ and $\gamma$. With a pre-existing setup and the pendulum roughly centered in its range of motion, this process typically takes about 20 minutes. Otherwise, it may take a couple hours. 

Data collection involved setting up the parameters in the function generator or Arduino, and then simply letting the experiment run. Since $f_0\approx 10 \,\mathrm{mHz}$, we typically used a sampling frequency $f_s=1 \mathrm{Hz}$. Then a reasonable dataset can be collected in $50/f_0$ seconds, under 90 minutes. Alternatively, data collection could run overnight, perhaps varying the duty cycle to collect several sequential datasets. The manufacturer-supplied software is limited to save the most recent 9 hours of data at $f_s=1$ Hz, or 18 hours at $f_s=0.5$ Hz. The results shown here use $f_s=1$ Hz. 
\section{Results and analysis}
We collected a wide range of data, representing different targets and optical setups. Here, we will show the best data, which was collected during the night, when the building was quiet; during the winter, when the room temperature was more stable; and with the glass mirrors. However, we will briefly discuss some of our other results as well. 

\begin{figure}[h!]
\centering
\includegraphics[width=3in]{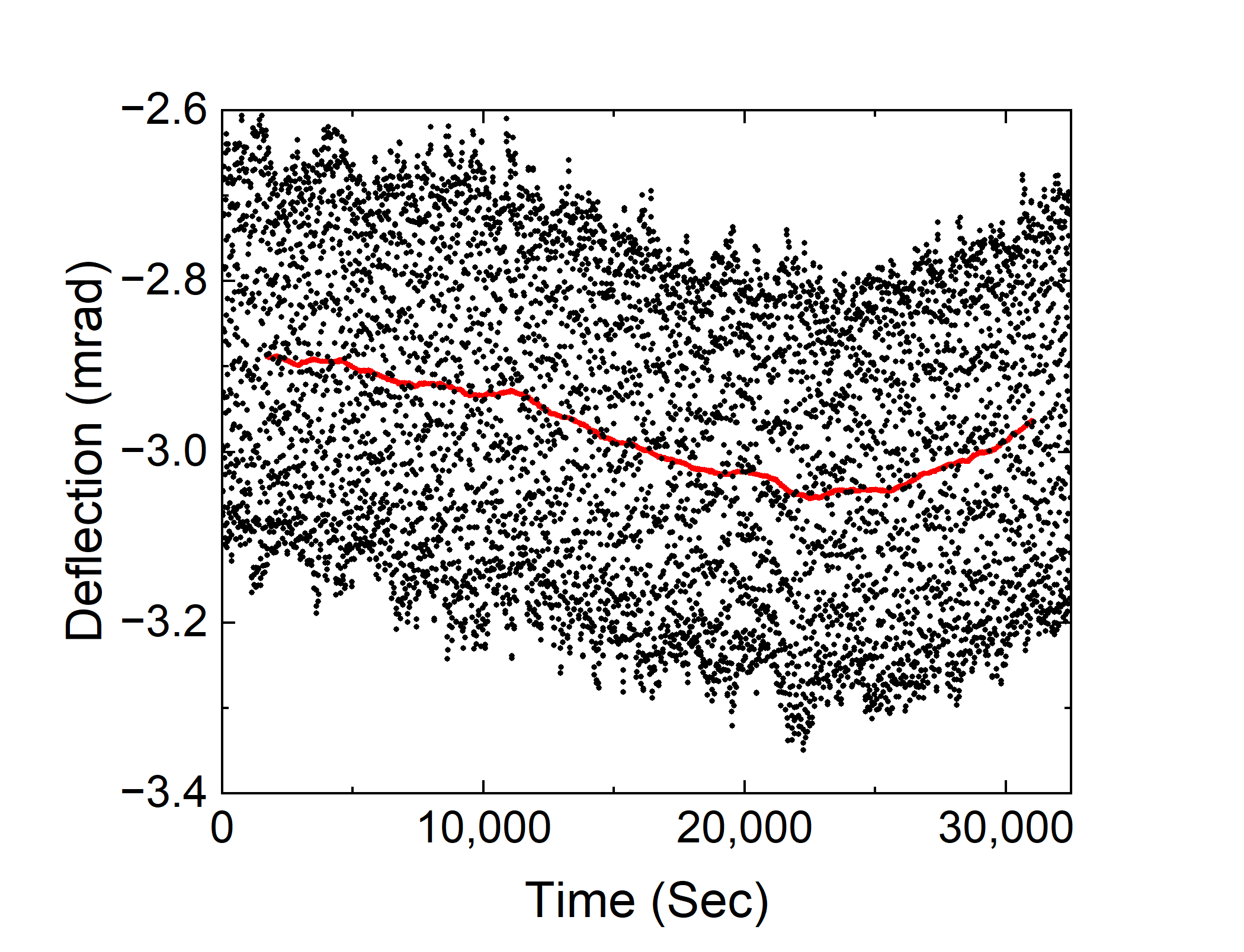}
\includegraphics[width=3in]{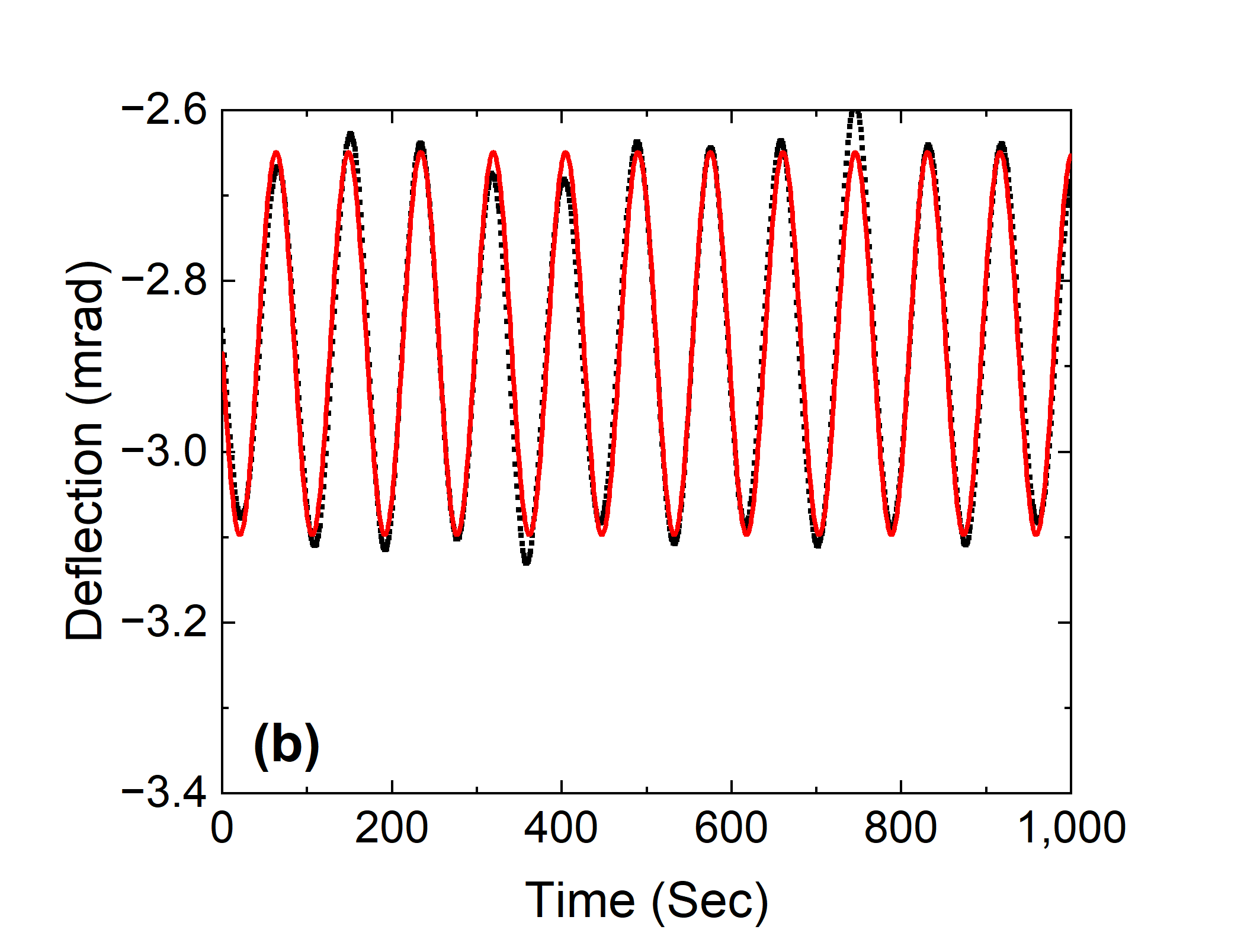}
\caption{Two approaches to analyzing the time-dependent pendulum motion.\textbf{(a)} Angular deflection over time, showing typical drift (1/5 datapoints shown for clarity). These data yield an RMS deviation of $0.17\pm0.01$ mrad-rms from the mean (red line, running average over 20 periods), corresponding to an amplitude of $0.24\pm0.01$ mrad. \textbf{(b)} A short section of this data with a sine fit yielding $A=0.22\pm0.01$ mrad. }
\label{timeanal}
\end{figure}

When the optical power was sufficiently large, we found that the time-dependent pendulum motion could be easily analyzed. As shown in Fig. \ref{timeanal}, the amplitude can be qualitatively estimated by eye, or quantitatively analyzed by either fitting the time series data to a sine function, or by calculating the RMS divergence of each datapoint from its mean value. Because there is some drift in the signal, we calculated the mean as a running average over 20 periods (1706 s). The RMS result is then multiplied by $\sqrt2$ to give the corresponding peak amplitude. 

\begin{figure}[h!]
\centering
\includegraphics[width=3in]{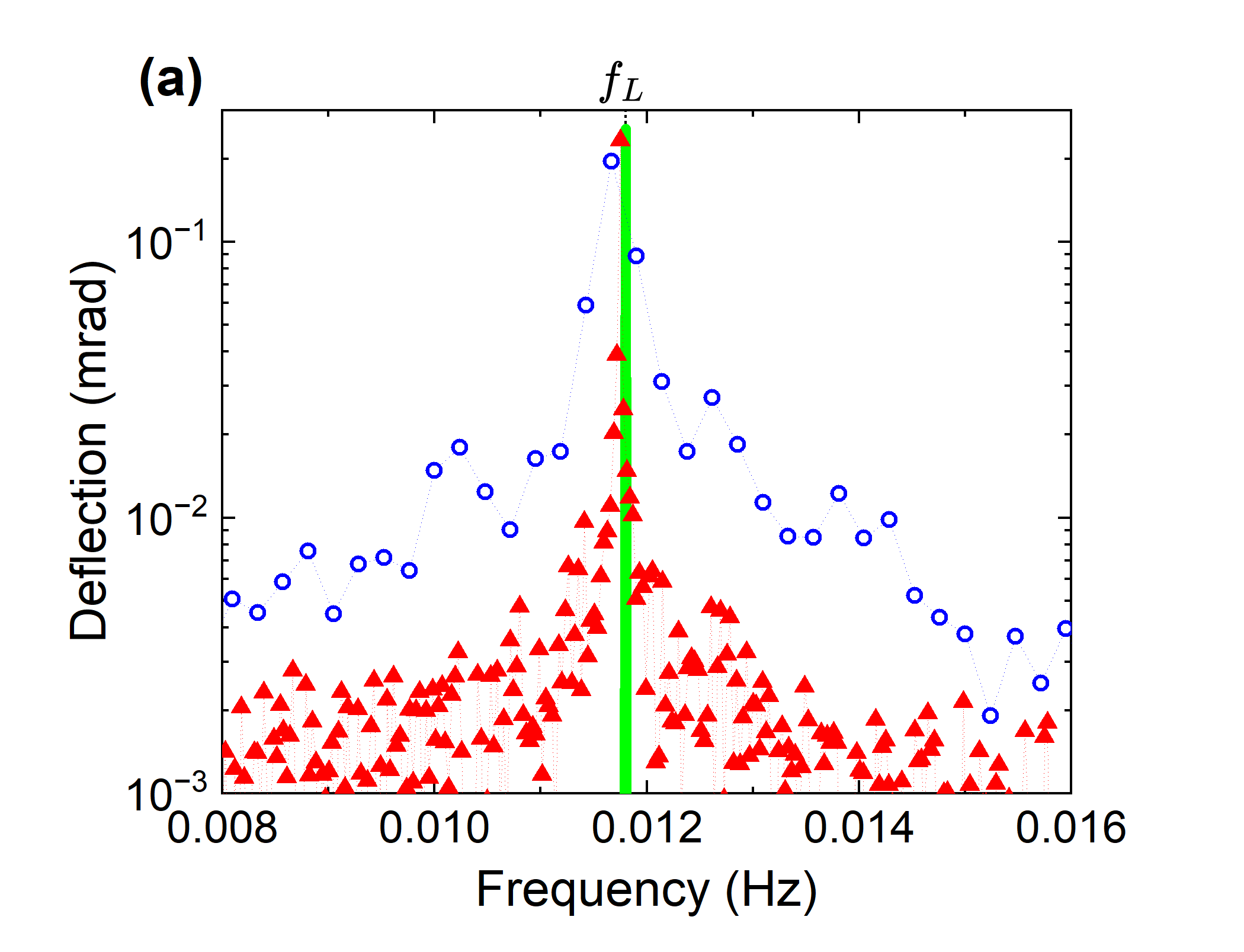}
\includegraphics[width=3in]{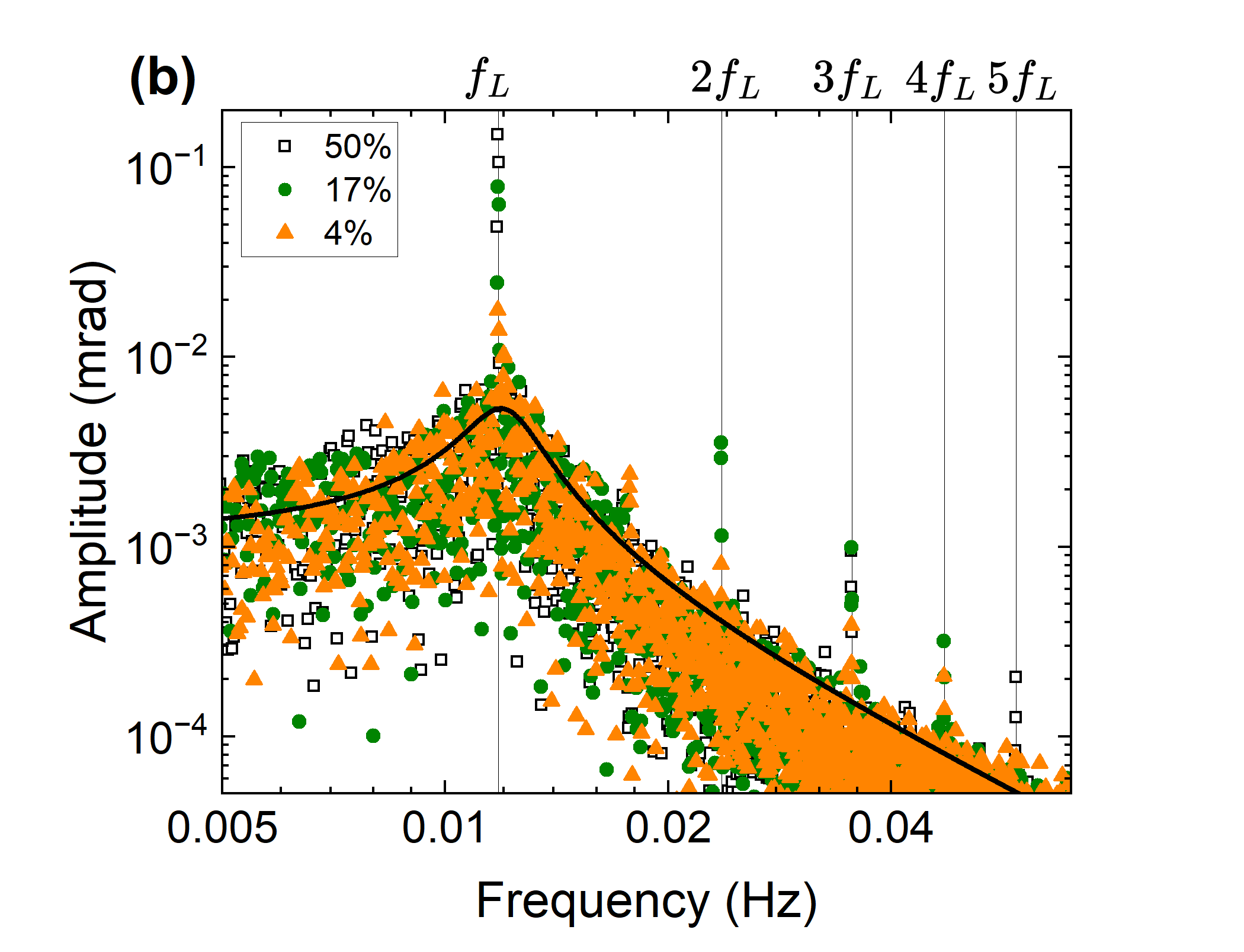}
\caption{Frequency-dependence of the driven pendulum motion. \textbf{(a)} The DFT, using a rectangular window, of the data shown in Fig. \ref{timeanal}, yielding $A=0.24$ mrad at $f=f_L$ for both a 4200 sec subset (open circles) and the full 9 hours of data (triangles). The power contained in the peaks is the same despite differences in linewidth and noise. The theoretical peak at $f_L$ is shown in the limit of zero peak width (green line). \textbf{(b)} Response of the pendulum illuminated by two laser diodes, with varying duty cycles as labeled; these are 9 hour datasets analyzed with a Hanning window to better show the weak, off-resonance signals and noise. Note the 50\% duty cycle signal lacks peaks at even multiples of $f_L$, as expected. The background noise has the same shape as $B(f)$ (see Eq. (\ref{B}) and (\ref{Bnoise})) indicating white noise (solid line, equal to $K\tilde{\phi}_B(f)$ with $K=3.5\times 10^4$).}
\label{DFT}
\end{figure}

Weaker driving signals can be observed using the Discrete Fourier transform (DFT) as shown in Fig. \ref{DFT}, and this additionally gives the full frequency-dependent response of the pendulum. These data are analyzed using the Fast Fourier Transform algorithm in OriginLab\cite{Origin}, with a rectangular window for quantitative results. Window (or, apodization) functions are helpful in DFT analysis because the dataset is not infinitely long, which can introduce nonidealities into the results known as spectral leakage. A rectangular window simply makes the beginning and end datapoints zero. It gives the most quantitative results; other window functions can be used to reduce noise and narrow the line widths, but tend to suppress the peak magnitudes as well. We used a Hanning window to visualize weaker off-resonance contributions to the spectra. As long as datasets were composed of 40 or more oscillation periods, spectral leakage wasn’t an issue. These topics are discussed a bit more in the supplementary materials.

The DFT spectra shown here each result from a single long dataset. This improves the frequency resolution of the DFT and reduces the magnitude of the background noise (proportional to $n^{-1/2}$ where $n$ is the length of the dataset). The background noise is stochastic, so to better see the functional form of the \textit{noise}, one can average multiple datasets; such averages are not shown here. 

We represent all our DFT data in amplitude form (as opposed to power or rms values) for easy visualization and comparison with the time-series data. The DFT peaks sometimes consist of a few datapoints. Labeling each data point in the peak as $A_i$, for $N$ such data points the total deflection $A$ corresponding to a single peak was calculated as
\begin{equation}
A=\sqrt{\sum_{i=1}^N {A_i}^2}.
\label{amplitude}
\end{equation}
This procedure can also be described as finding the total power in the peak. The data shown in Figs. \ref{timeanal} and \ref{DFT}a are all from the same dataset, representing driving the glass mirrors with four laser diodes for a combined $20\pm0.2$ mW of power, with square wave modulation at $f_L=f_0=11.8$ mHz and $d=0.5$. The amplitude of $\phi(t)$ is measured in the three ways described above, with consistent results. A sine fit gives $A=0.22\pm0.01$ mrad, and using the RMS deviation from the mean, multiplied by $\sqrt2$, results in amplitudes of $0.24\pm0.01 \, \mathrm{mrad}$. Analyzing the DFT using Eq. \ref{amplitude} yields a total amplitude $0.24\pm0.01$ mrad at $f=f_L$, with uncertainty corresponding to the noise floor. The predicted deflection is $0.26$ mrad.

In Fig. \ref{DFT}b, the DFT results are shown for several different duty cycles. These show additional peaks, which agree well with theoretical predictions. From Eq. (\ref{phiBtau}), the pendulum’s response to the square wave modulated light has peaks at discrete frequencies
\begin{equation}
|\tilde{\phi}(f)| =
\begin{cases}
	\tau_nB(f) & f=nf_L \, , \\
	\tau_{\mathrm{noise}}B(f) &  \mathrm{else} \, .
\end{cases}
\label{25}
\end{equation}
where $\tau_{\mathrm{noise}}$ is the torque resulting from any background noise.  The $\tau_n$ values are related to the total incident  optical  force $F_n$ and the optical power $P_L$ by Eq. (\ref{optF}), giving
\begin{equation}
\tau_n=F_n r=\frac{P_{n,\mathrm{eff}}}{c}r
\end{equation}
where $P_{n,\mathrm{eff}}$ represents the actual amplitude of the optical power interacting with the mirror at frequency $nf_L$. From Eq. (\ref{FTsq}), we have
\begin{equation}
P_{n,\mathrm{eff}}=GP_L\left| \frac{2}{n\pi} \sin(dn\pi) \right|
\label{Peff}
\end{equation}
where $P_L$ is the total driving laser power incident on the windows. The constant $G$ represents the effects of the optical path. We use 
$G = (1 + \alpha) T [ 1 + (1-T) ] $,
where $T$ is the optical transmission of the enclosure window, and the $(1-T)$ term represents retro-reflected light from the glass that hits the mirror a second time. As expected for glass, we measured $T=0.89\pm0.01$ by comparing incident, reflected, and transmitted light using the Newport power meter. Reflected light from the mirror gave $\alpha=0.6\pm0.1$  This yields $G=1.6$. However, the low value of $\alpha$ suggests diffusive reflection will also play a role, and we estimate this could cause our value of $G$ to be too low by as much as 5\%. 

The deflection at $f\ne nf_L$ has the same functional form as $B(f)$, indicating that we do have white background noise. The theoretical curve $K\tilde{\phi}_B$ is shown in Fig. \ref{DFT}b, where $K$ is a constant adjustable parameter; there are no other fitting parameters. The value of $K$ is typically 4-5 orders of magnitude larger than the predicted Brownian motion, suggesting that interactions with the surrounding air are not the primary driver of this noise signal. This background noise remains the same when there is no optical forcing.

One interesting result is that the signal to noise ratio in the DFT will be similar at any frequency; even though the magnitude of the pendulum oscillation decreases for $f_L>f_0$, the noise background decreases at the same rate. So, as long as DFT analysis is used, then a sensitive measurement can still be achieved with $f_L \ne f_0$. 

\begin{figure}[h!]
\centering
\includegraphics[width=3in]{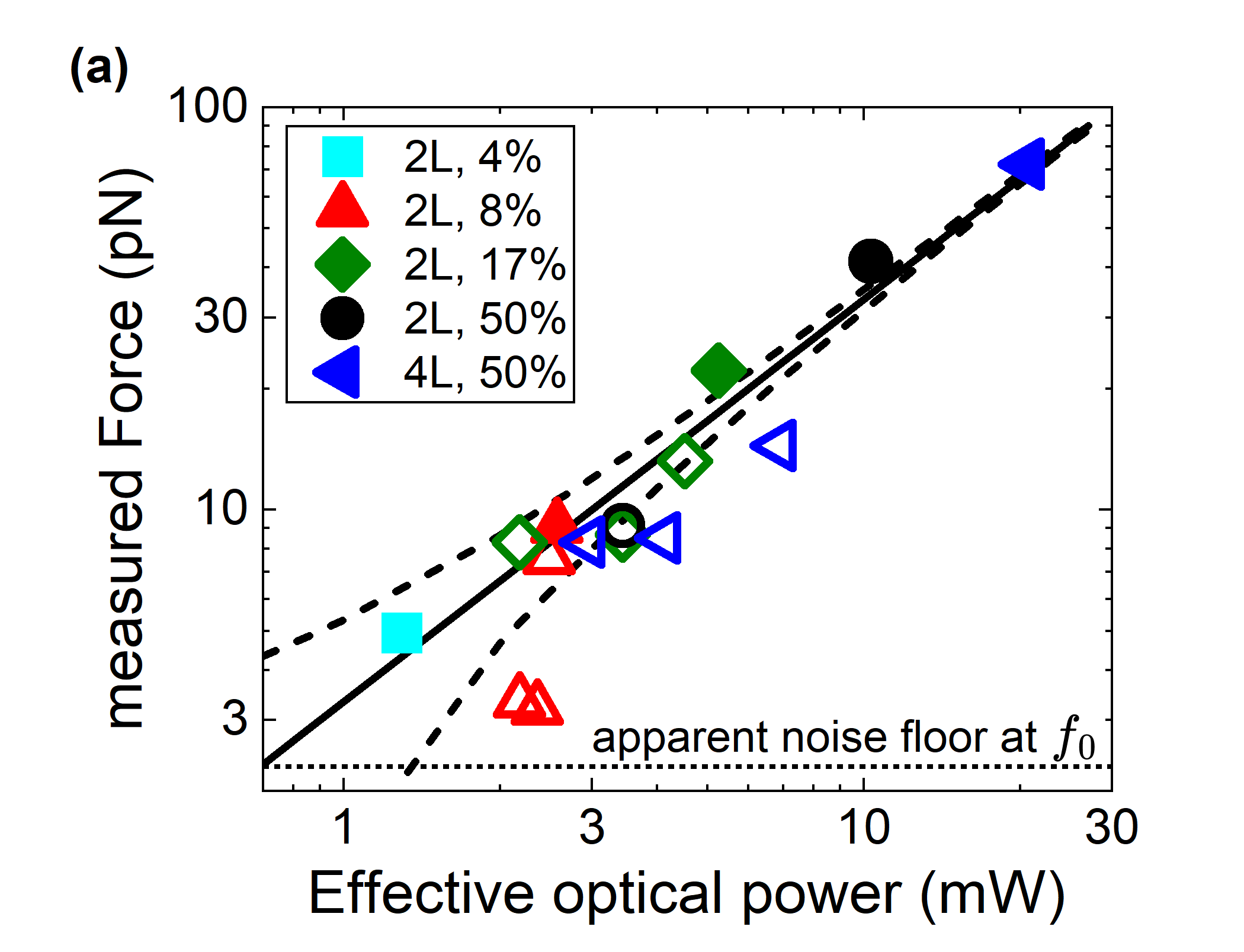}
\includegraphics[width=3in]{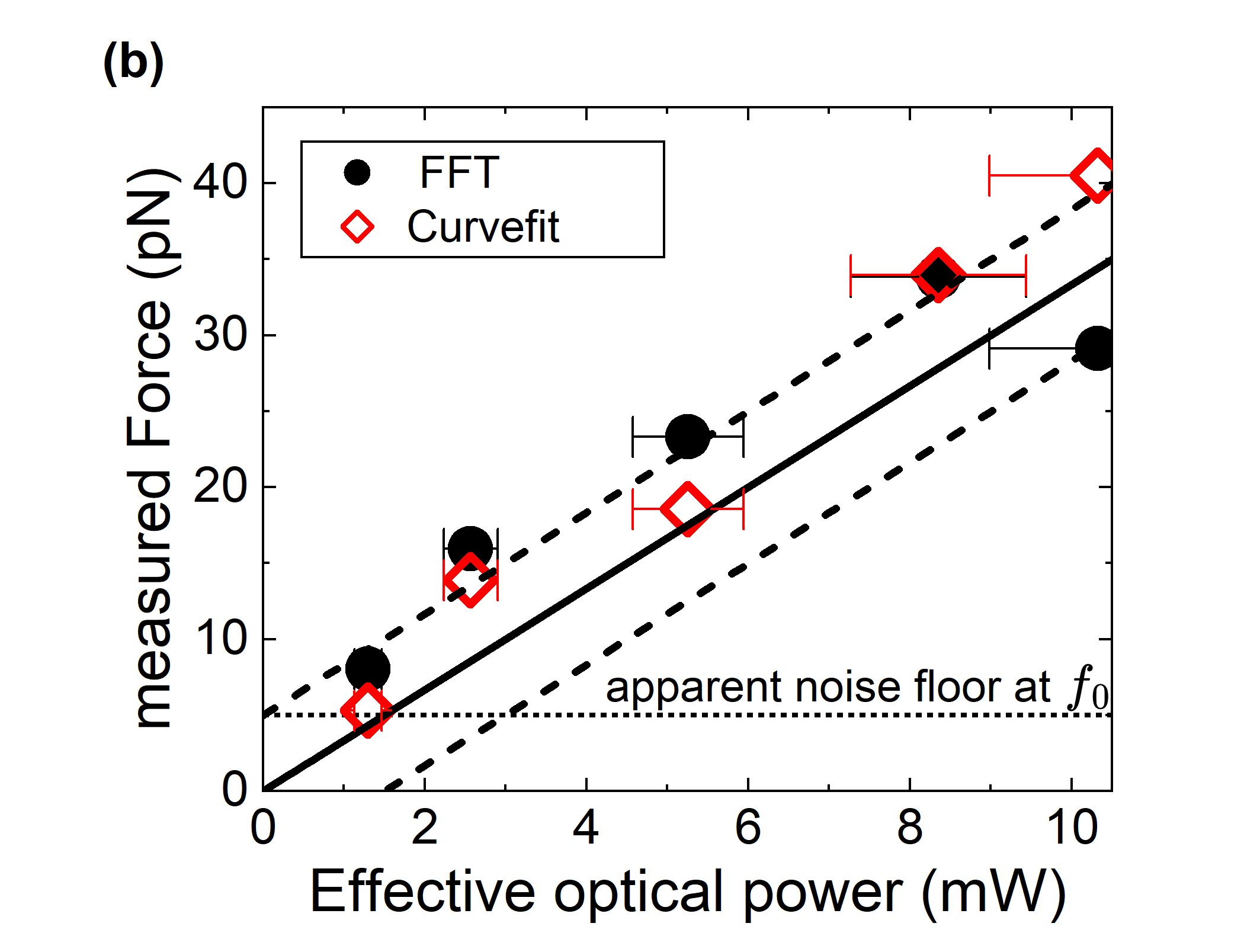}
\caption{Measured force obtained from $|\tilde{\phi}_f|$ using Eq. \ref{25}, as a function of the effective optical power, Eq. \ref{Peff}. In both figures, the solid line has a slope of $1/c$, while dashed lines show the predicted noise envelope. \textbf{(a)} Results from 9 hour datasets using four (4L) or two (2L) 5 mW laser diodes, pulsed at $f_0$ with the indicated duty cycles. Solid symbols are the primary Fourier component, $n=1$, and corresponding open symbols are for $n>1$; size corresponds to about 10\% uncertainty. \textbf{(b)} Result from a single overnight run of 5000 sec datasets at varying duty cycles, using two laser diodes. The same dataset was analyzed with both the DFT and with a sinewave curvefit to the central 1000 sec of data.}
\label{summary}
\end{figure}

The results of multiple measurements are summarized in Fig. \ref{summary}. These data are all analyzed with an FFT using a rectangular window for more quantitative results. Here, the symbol size corresponds to about 10\% of the signal to indicate the uncertainty, which was almost entirely systematic. The uncertainty in measured force originates primarily from values of $I$ and $r$. Uncertainty in $P_{\mathrm{eff}}$ is dominated by $G$, especially the exact role of multiple internal reflections and diffusive reflection. Statistical uncertainty in the force is not included in the error bars. Since it should be on the order of the measured background signal,  we’ve instead visualized it as a envelope surrounding the predicted result. Thus we expect $n=1$ data to fall between the dashed lines in the figure, where $n$ is the order of the driving frequency. The expected slope is simply $1/c$, shown as a solid line. The $n>1$ data are systematically smaller than expected. These are tiny components of the Fourier spectrum, and though its challenging to quantify their uncertainty, we're excited to see that they come close to the expected values.
The effective optical power was adjusted in three ways: using neutral density filters, adjusting the number of laser diodes that were turned on, and adjusting the duty cycle $d$. These approaches all work as expected; however, adjusting the duty cycle provides an elegant way to access the full range of possible optical powers with a simple electronic setting that can be easily automated, and those results are shown here. Once the pendulum is set up, it’s helpful not to come anywhere near the optical table as datapoints are taken so as to avoid any vibrations. 

\begin{figure}[tbh]
    \begin{minipage}{0.49\textwidth}
            \includegraphics[width=3in]{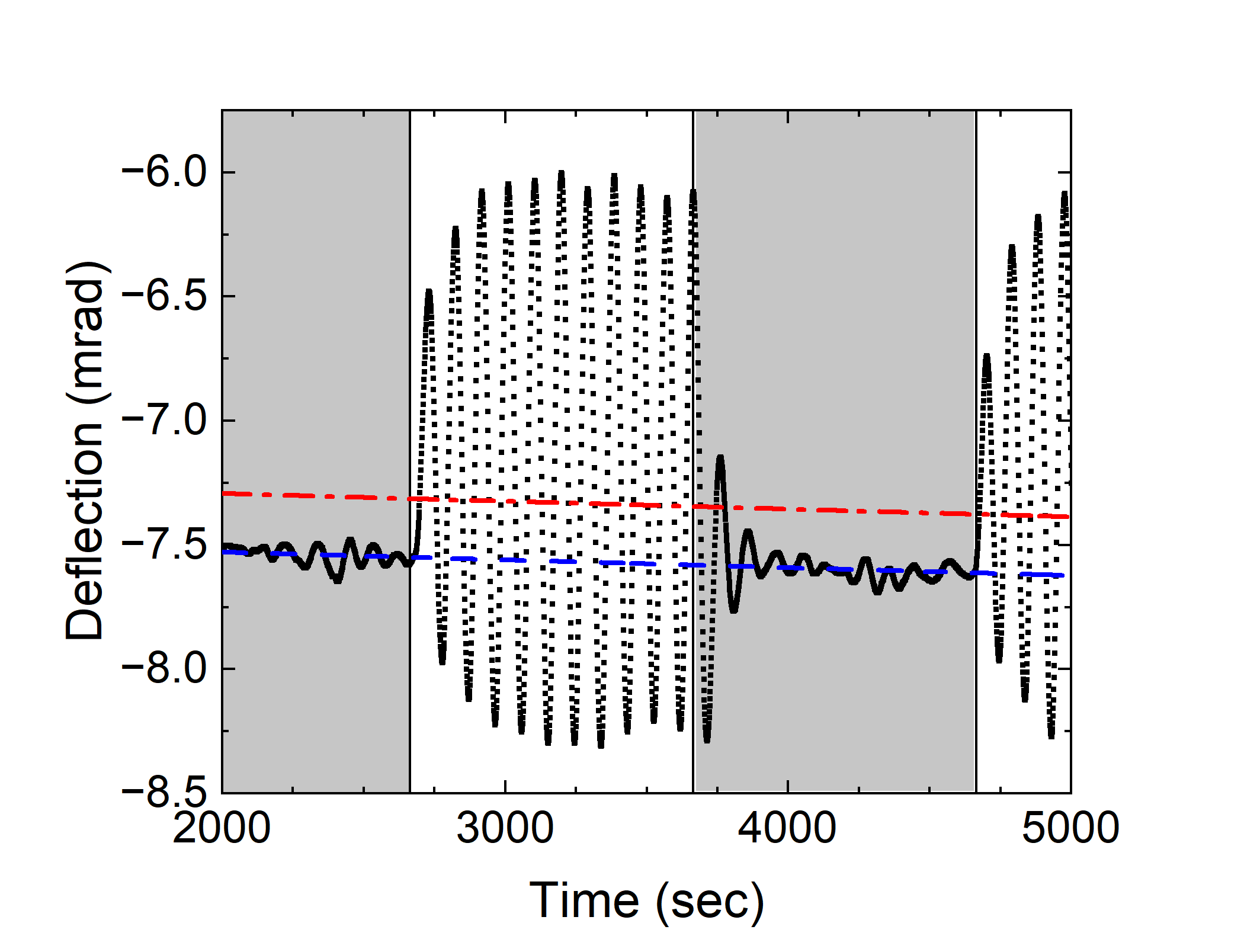}
    \end{minipage}%
    \hspace{\fill}%
    \begin{minipage}{0.48\linewidth}
        \caption{Pendulum with foil reflectors, responding to 5mW excitation at $f_L=f_0$ and 50\% duty cycle. Regions of no illumination (grey shading) are included for comparison; red and blue lines are linear fits to the light-on and light-off regions respectively, showing the expected DC offset (because only one laser diode was used), as well as a slight drift.}
        \label{foil}
    \end{minipage}
\end{figure}

We also tested a foil reflector, as previously mentioned. This design had $Q=2.9$, $\kappa=1.76\times10^{-7}$ N/m, and $f_0=10.9$ mHz. Figure \ref{foil} shows its response to one 5 mW laser diode at $f_L=f_0$ and with $d=0.5$; the response was more than an order of magnitude greater than expected from the optical force alone; the higher response was probably due to heating. While this not ideal for measuring  optical  force, it does make the driven pendulum response easy to study; a solid black painted target also yields a very strong force. In Fig. \ref{foil}, one can clearly observe both the light-induced DC offset (because only one laser was used) and the AC response. Upon a change of illumination, the DC offset changes rapidly, but the AC pendulum oscillation shows the expected initial transient response, followed by steady-state behavior. 

Heat and drift both seem to affect the background noise. This was a problem during summer in our non-air-conditioned room, when temperatures varied dramatically over the day and night. By monitoring the optical lever signal over time, we confirmed that the pendulum position did remain steady even as the sensor drifted, so the observed drift was an electronic effect and not a real change in position. We’d recommend using a climate-controlled room to avoid this issue entirely. The data shown here were collected during winter, when the room maintained a relatively stable temperature. Some real drift also clearly occurs at all times, but at much smaller magnitudes.

\section{Conclusions}
These results demonstrate that the  optical  force from an everyday light source can be measured using a commercially obtained torsion pendulum with built-in capacitance sensor, operating in air. Results are consistent with theoretical expectations, indicating that glass reflectors sufficiently prevent differential heating of the reflector surface, allowing the  optical  force to dominate. Measurement uncertainty is dominated by the optical path, both the placement of the mirrors and laser spots, and the subsequent specular and diffusive reflections; this could be improved in future studies. 

Targets that do seem to show convective forcing cause a much larger oscillator response, and these can also be valuable for exploring, generally, the forced damped oscillator. Drawings and files to 3D print mounts for Mylar or foil targets are included in the supplemental information.  

A compelling short sophomore-level lab might employ an instructor-setup pendulum and optics, using a single illumination power, and allowing students to observe the laser-induced deflection, comparing the amplitude of the motion to expectations. When using four lasers at $d=0.5$, the pendulum response is strong and students should be able to collect sufficient data in a short amount of time, similar to Fig. \ref{timeanal}. Expanding the time frame of the lab to several sessions would allow students to collect data overnight, collect multiple datapoints, explore the Fourier transform, and/or analyze the noise. 

More advanced students could use this setup in a multi-week to semester-long lab experience and could choose from several pathways to create their own project. If the pendulum itself isn’t already set up, then determining the experimental details, implementing the design, and characterizing the pendulum’s behavior can itself be an interesting undertaking. From there, students could choose the direction of their research project. Here we have oriented the discussion toward measurement of the optical force, however, it is also possible to use this same approach and the known force to characterize the pendulum response function by measuring its deflection at a series of driving frequencies. 

These results also suggest more substantial projects, as there are unanswered questions here including the exact origin of noise in the system, and the role of of the capacitive position sensor in energy dissipation. There would also be opportunities to create experiments using position as active feedback to adjust the laser driving force. Indeed, we believe the most exciting aspect of this setup is its flexibility as a platform for student-designed projects. This work just scratches the surface of what would be possible. 

\section{Acknowledgements}
The authors gratefully acknowledge Jay Ewing for assistance with design and 3D printing, and Darrell Schroeter for helpful feedback. Thanks to Leo Grossman, Alex West, Ella Morton, Laura Estridge, and Noah Miller for their participation and insights. This work was partly supported by the Delord-Mockett Student Research Fund.

\section{Author Declarations}
The authors have no conflicts to disclose.

\end{document}